\documentclass[jkps,preprint,showpacs,showkeys, numbers]{revtex4}
\usepackage{graphicx}
\usepackage{amssymb}
\usepackage{amsmath}
\usepackage{bm}

\usepackage{graphicx}
\usepackage[export]{adjustbox}
\usepackage[utf8]{inputenc}
\usepackage[T1]{fontenc}
\usepackage{graphicx}
\usepackage[hidelinks]{hyperref}
\usepackage{microtype}
\usepackage{array}

\usepackage{rotating}

\usepackage{multirow}
\usepackage{float}

\usepackage{booktabs}

\usepackage{adjustbox}

\usepackage{graphicx}

\usepackage{verbatim}

\usepackage{color,soul}

\begin{document}
\setcounter{page}{0}
\title[]{Nonlinear Evidence of Investor Heterogeneity: Retail Cash Flows as Drivers of Market Dynamics}
\author{Gabjin \surname{Oh}}
\email{phecogjoh@chosun.ac.kr}
\affiliation{College of Business, Chosun University, Gwangju
61452}

\date[]{}

\begin{abstract}
    
This study measures the long memory of investor-segregated cash flows within the Korean equity market from 2015 to 2024. Applying detrended fluctuation analysis (DFA) to BUY, SELL, and NET aggregates, we estimate the Hurst exponent ($H$) using both a static specification and a 250-day rolling window. All series exhibit heavy tails, with complementary cumulative distribution exponents ranging from approximately 2 to 3. As a control, time-shuffled series yield $H \approx 0.5$, confirming that the observed persistence originates from the temporal structure rather than the distributional shape. Our analysis documents long-range dependence and reveals a clear ranking of persistence across investor types. Persistence is strongest for retail BUY and SELL flows, intermediate for institutional flows, and lowest for foreign investor flows. For NET flows, however, this persistence diminishes for retail and institutional investors but remains elevated for foreign investors. The rolling $H$ exhibits clear regime sensitivity, with significant level shifts occurring around key events: the 2018--2019 tariff episode, the COVID-19 pandemic, and the period of disinflation from November 2022 to October 2024. Furthermore, regressions of daily volatility on the rolling $H$ produce positive and statistically significant coefficients for most investor groups. Notably, the $H$ of retail NET flows demonstrates predictive power for future volatility, a characteristic not found in institutional NET flows. These findings challenge the canonical noise-trader versus informed-trader dichotomy, offering a model-light, replicable diagnostic for assessing investor persistence and its regime shifts.

\end{abstract}

\pacs{05.45.Df, 05-10-a, 05.40.Fb, 05.90+m}

\keywords{Econophysics, Heterogeneity, Retail investor, Market efficiency}

\maketitle

\section{Introduction}

Financial markets are complex systems comprising heterogeneous investors. This paper tests the heterogeneous-investor hypothesis by examining whether the trading flows of three distinct groups---retail, institutional, and foreign---exhibit unique dynamics, and specifically, whether retail activity can be characterized as noise. Our analysis focuses on the daily buy, sell, and net-buy flows for each group within the Korean equity market. These three groups exhibit distinct practical behaviors: retail investors typically trade in numerous small orders with short horizons; institutional investors operate under portfolio and benchmark constraints while managing inventory at scale; and foreign investors integrate cross-market information, often executing trades programmatically across different time zones. To detect persistence and its corresponding regime changes without imposing strong modeling assumptions, we employ detrended fluctuation analysis (DFA) to compute both static and rolling Hurst exponents \cite{Serinaldi2010, Barunik2010, Yin2013,Teng2018,Peng1994}. According to the noise trader hypothesis, correlations should decay rapidly, yielding a Hurst exponent of $H \approx 0.5$; systematic deviations from this value ($H > 0.5$) indicate a persistent temporal structure. Consequently, our research design focuses on determining whether each group’s flows exhibit long-range dependence and how this dependence varies across the buy, sell, and net-buy series. We then link this evidence to complex-systems perspectives that extend beyond the traditional homogeneous representative agent model, aligning with the fractal market hypothesis \cite{Peters1994, Horta2014,Blackledge2022,Dar2017,Caraiani2012,Vogl2023}.

Most empirical tests of scaling and persistence in finance have relied on prices and volumes. Such proxies mix the actions of heterogeneous agents and are endogenous to microstructure frictions (e.g., bid--ask bounce, inventory control, volatility regimes), which limits their power to discriminate between the Efficient Market Hypothesis (EMH) and the Fractal Market Hypothesis (FMH) \cite{Peters1994, Fama1970,Lo1991,Kristoufek2013,Huang2019,Bariviera2017}. Our research design overcomes these limitations by using investor-segmented cash-flow series: daily buy, sell, and net-buy amounts reported separately for retail, institutional, and foreign investors. These labeled flows allow for like-for-like comparisons across investor types, isolate heterogeneity at its source, and permit an explicit analysis of the net-buy series to assess the interaction between buying and selling pressures. Applying DFA directly to these flows (rather than to returns or raw volumes) reduces mechanical autocorrelation arising from price discreteness and seasonal trading patterns, keeps preprocessing to a minimum, and yields a direct, replicable measurement of long-range dependence at the level of specific investor types \cite{Qian2011,Gomez2022}. Therefore, this unique dataset transforms the long-standing debate over homogeneity versus heterogeneity into a measurable property of market microstructure, all within a framework that is consistent with EMH/FMH tests yet focused on agent-specific dynamics.

Our empirical strategy is straightforward and specifically designed to address the question of investor heterogeneity. We estimate DFA-based Hurst exponents for nine distinct flow series, denoted by $X^{(g,a)}_t$, where the investor group $g \in \{\text{retail, institutional, foreign}\}$ and the flow type $a \in \{\text{BUY, SELL, NET}\}$, with the net flow defined as $NET^{(g)}_t = BUY^{(g)}_t - SELL^{(g)}_t$. The analysis proceeds in two stages. First, we compute static estimates over the entire sample period to quantify persistence for each group and flow type; if retail activity were purely noise, its corresponding Hurst exponent ($H$) would be approximately 0.5. Second, we use rolling estimates based on a one-year overlapping window with a weekly step to track temporal variations and identify potential regime shifts, particularly during periods of market-wide stress. To ensure robustness, our statistical inference is anchored by a key benchmark: time-shuffled series, which destroy the temporal structure while preserving the marginal distribution of the flows \cite{Kristoufek2014,Vogl2023FRL}.

We investigate three issues central to investor heterogeneity and crisis response: (i) whether buy and sell flows for each group exhibit long–range dependence; (ii) whether the net–buy series (BUY--SELL) weakens or preserves that dependence, and how this differs across groups; and (iii) how persistence varies over time and across investor types, with special attention
to market–wide stress episodes such as the COVID–19 shock \cite{Wu2020,Lim2022}. Empirically, we estimate static and rolling Hurst exponents with detrended fluctuation analysis (DFA) and benchmark measurements against shuffled and phase–randomized series. We define pre–shock, shock, and post–shock windows to quantify level shifts and volatility in $H_t$ and to test for synchronization or divergence across groups during crises. The paper’s substantive contribution is to shift tests of heterogeneity from prices and raw volumes to investor–segregated cash–flow data, enabling direct, like–for–like comparisons across investor types and flow directions and providing evidence on how retail, institutional, and foreign investors adjust their dynamics under crisis conditions \cite{Li2010,Kelley2013,Horta2016,Oh2017,Tang2021,Da2019,Karaomer2022}.

The remainder of the paper is organized as follows: Section II details the data and methodology employed in our analysis. Section III presents the findings of our study. Finally, Section IV offers conclusions drawn from our research.

\section{Data and methodology}

We use daily \emph{trading amounts} (cash-flow values, KRW) by investor type from \textsc{FnGuide} for 2015–2024. For each listed firm \(i\), trading day \(t\), and investor group
\(g\in\{\text{retail},\text{institutional},\text{foreign}\}\), the data report \(\mathrm{BUY}^{(g)}_{i,t}\) and \(\mathrm{SELL}^{(g)}_{i,t}\).
To study group-level dynamics on a common calendar, we aggregate across firms each day:
\begin{equation}
\mathrm{BUY}^{(g)}_{t}=\sum_{i=1}^{N_t}\mathrm{BUY}^{(g)}_{i,t},\qquad
\mathrm{SELL}^{(g)}_{t}=\sum_{i=1}^{N_t}\mathrm{SELL}^{(g)}_{i,t},
\label{eq:agg}
\end{equation}
where \(N_t\) is the number of firms trading on day \(t\).
We define \emph{net-buy} as the difference
\begin{equation}
\mathrm{NET}^{(g)}_{t} \equiv \mathrm{BUY}^{(g)}_{t}-\mathrm{SELL}^{(g)}_{t}.
\label{eq:net}
\end{equation}
The analysis therefore considers nine market-wide series
\[
X^{(g,a)}_t \in \Big\{\mathrm{BUY}^{(g)}_t,\ \mathrm{SELL}^{(g)}_t,\ \mathrm{NET}^{(g)}_t\Big\}
\quad\text{for}\quad g\in\{\text{retail},\text{institutional},\text{foreign}\}.
\]
We retain native units, align all groups to the exchange trading calendar, and exclude non-trading days. 
Our objective is to measure temporal persistence in investor–segregated trading amounts and to evaluate how that persistence differs across groups and over time. We use \emph{detrended fluctuation analysis} (DFA) as introduced by Peng \emph{et al.}~\cite{Peng1994}. DFA yields a scale–robust estimate of the Hurst exponent \(H\) without imposing a specific stochastic model.

%\subsection{Notation and series}
%For each investor group \(g\in\{\mathrm{retail},\mathrm{institutional},\mathrm{foreign}\}\) and trading %day \(t=1,\dots,T\), let
%\[
%\mathrm{BUY}^{(g)}_{t},\qquad \mathrm{SELL}^{(g)}_{t}
%\]
%denote market–wide cash–flow amounts aggregated across firms on day \(t\) (see Eq.~\eqref{eq:agg} in %the Data section). We define the \emph{net–buy} series as
%\begin{equation}
%\mathrm{NET}^{(g)}_{t}\equiv \mathrm{BUY}^{(g)}_{t}-\mathrm{SELL}^{(g)}_{t}.
%\label{eq:net_series}
%\end{equation}
%The nine analysis series are \(X^{(g,a)}_t\in\%{\mathrm{BUY}^{(g)}_t,\mathrm{SELL}^{(g)}_t,\mathrm{NET}^{(g)}_t\}\).
%Baseline estimations use the native units; robustness checks consider a sign–preserving transform for the net series,
%\begin{equation}
%\tilde X^{(g,\mathrm{NET})}_t=\mathrm{sign}\!\big(X^{(g,\mathrm{NET})}_t\big)\,\log\!\big(1+\lvert X^{(g,\mathrm{NET})}_t\rvert\big),
%\label{eq:signlog}
%\end{equation}
%and mild winsorization of extreme outliers. These choices do not alter the qualitative results and are reported with the Results.

\subsection{Detrended fluctuation analysis (static \(H\))}
The nine analysis series are \(X^{(g,a)}_t\in\{\mathrm{BUY}^{(g)}_t,\mathrm{SELL}^{(g)}_t,\mathrm{NET}^{(g)}_t\}\). Given a mean–adjusted series \(X_t\) of length \(T\), construct the integrated profile
\begin{equation}
Y(\tau)=\sum_{t=1}^{\tau}\big(X_t-\bar X\big),\qquad \tau=1,\dots,T.
\label{eq:profile}
\end{equation}
Choose a \emph{scale grid} \(\mathcal N=\{n_1,\dots,n_K\}\) with logarithmic spacing. For each \(n\in\mathcal N\), partition \(Y\) into \(M(n)=\lfloor T/n\rfloor\) non–overlapping blocks of length \(n\). In each block \(b=1,\dots,M(n)\), fit and remove a polynomial trend of order \(m\) to obtain residuals \(r_{b,j}^{(n)}\) for \(j=1,\dots,n\). The block RMS is
\[
F_b^2(n)=\frac{1}{n}\sum_{j=1}^{n}\big(r_{b,j}^{(n)}\big)^2,
\]
and the fluctuation function at scale \(n\) is the block average
\begin{equation}
F(n)=\sqrt{\frac{1}{M(n)}\sum_{b=1}^{M(n)}F_b^2(n)}.
\label{eq:Fdef}
\end{equation}
For long–range dependent processes one expects
\begin{equation}
F(n)\propto n^{H}\quad\Longleftrightarrow\quad
\log F(n)=a+H\,\log n+\varepsilon_n.
\label{eq:scaling}
\end{equation}
We estimate \(H\) as the OLS slope in Eq.~\eqref{eq:scaling} on an \emph{admissible} subset \(\mathcal N^\star\subset\mathcal N\). Very small scales (detrending unreliable) and very large scales (few blocks) are excluded \emph{ex ante} and kept identical across series. Unless otherwise stated, we use DFA(2) (\(m=2\)), and verify stability under DFA(1).

%\paragraph*{Scale diagnostics.}
%We report the number of blocks \(M(n)\), the regression \(R^2\), and \emph{local slopes}
%\[
%H_{\mathrm{loc}}(n)=\frac{\Delta\log F(n)}{\Delta\log n}
%\]
%computed on three–point neighborhoods to check approximate scale–invariance within \(\mathcal N^\star\). As a leverage check, we re–estimate after dropping the smallest and largest scales in \(\mathcal N^\star\).

\subsection{Rolling DFA (time variation)}
To track time variation and responses to market stress, we compute \emph{rolling} Hurst exponents. We fix a window length \(W\) of 250 trading days and a step size \(s\) of 5 trading days. For a window starting at \(t_w \in \{1, 1+s, 1+2s, \dots\}\) with \(t_w+W-1 \le T\), we define the index set
\[
\mathcal{T}_w = \{t_w, \dots, t_w+W-1\}.
\]
We then apply the static DFA procedure to the subset \(X_{\mathcal{T}_w} = \{X_t : t \in \mathcal{T}_w\}\) using the same scale range \(\mathcal{N}^\star\) and detrending order \(m\), and record the resulting exponent \(H_w\). This exponent is assigned to the end point of the window, \(t_w + W - 1\), which yields a time series \(\{H_w^{(g,a)}\}\) for each group and flow type. For the crisis analysis (e.g., COVID-19), we summarize the levels and volatilities of \(H_w\) within calendar windows defined in the Results section. 

% --- refs for methodology (add to .bib) ---
% \bibitem{Peng1994} C.-K.~Peng, S.V.~Buldyrev, S.~Havlin, M.~Simons, H.E.~Stanley, and A.L.~Goldberger,
% ``Mosaic organization of DNA nucleotides,'' Phys.\ Rev.\ E \textbf{49}, 1685--1689 (1994).
% \bibitem{Peng1995} C.-K.~Peng, S.V.~Buldyrev, S.~Havlin, J.M.~Hausdorff, J.E.~Mietus, H.E.~Stanley, and A.L.~Goldberger,
% ``Fractal mechanisms and heart rate dynamics,'' Chaos \textbf{5}, 82--87 (1995).

\section{Results}

\begin{figure}[!ht]
\begin{center}
\includegraphics[width=14cm,height=14cm,keepaspectratio]{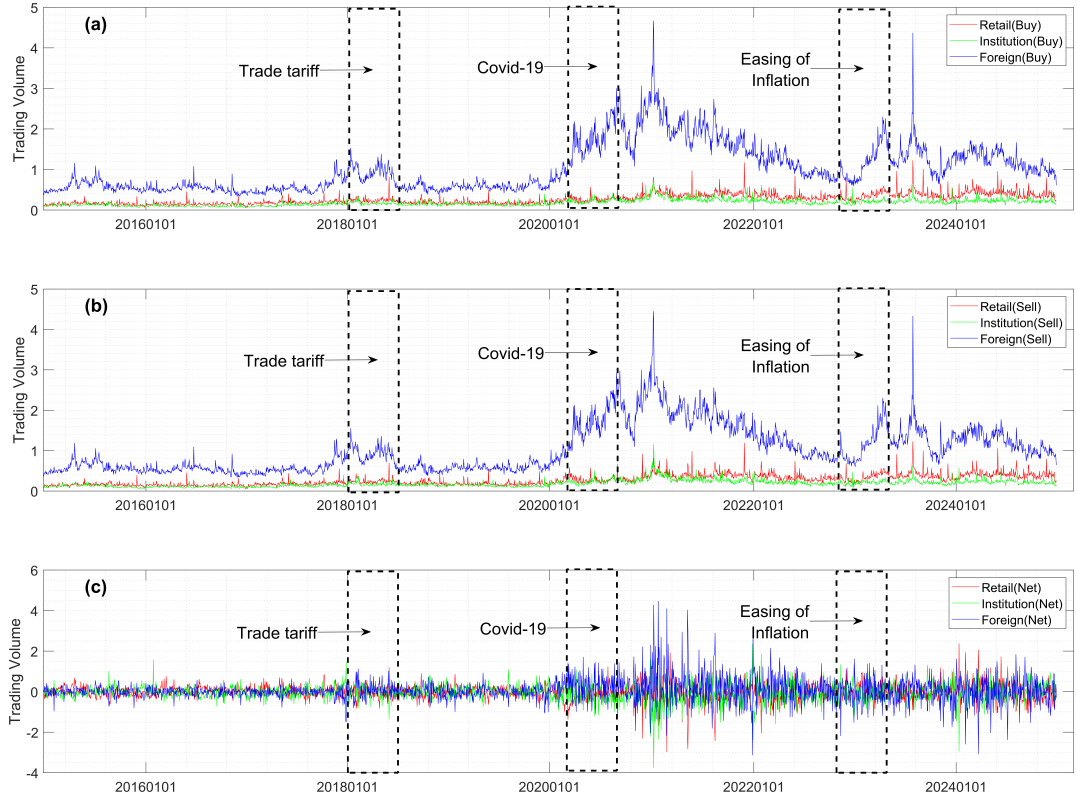}
\end{center}

\caption[Investor-segregated trading volumes, 2015--2024]{Daily investor-segregated trading volumes in the Korean equity market from 2015 to 2024. The panels display time series for three investor groups: retail (red), institutional (green), and foreign (blue). Panel~(a) shows total BUY volumes, (b) total SELL volumes, and (c) the NET (BUY$-$SELL) flows. Vertical dashed boxes highlight three key periods of analysis: the 2018--2019 trade-tariff tensions, the COVID-19 episode beginning in early 2020, and the 2022--2024 disinflation phase. The y-axis represents the daily cash-flow volume.}
\label{fig:rawflows}
\end{figure}

We begin with a descriptive overview of the investor-segregated flows to provide context for the subsequent inference (see Fig.~\ref{fig:rawflows}). Our analysis then proceeds in two main stages: first, we quantify persistence using static Hurst exponents derived from DFA, and second, we examine time variations using rolling Hurst exponents. The crisis periods analyzed in this study---specifically, the 2018--2019 trade-tariff tensions, the COVID-19 episode beginning in 2020, and the 2022--2024 disinflation phase---are highlighted by dashed boxes in Fig.~\ref{fig:rawflows}.

As shown in Panels (a) and (b) of Fig.~\ref{fig:rawflows}, retail buy and sell volumes dominate the market-wide flow and exhibit pronounced volatility clustering, with large bursts aligning with the highlighted crisis periods. While institutional and foreign flows are smaller in magnitude, they display episodic spikes, particularly around the onset of the COVID-19 pandemic. This asymmetry in trading volume across groups is a persistent feature throughout the sample period. Panel (c) of Fig.~\ref{fig:rawflows} plots the net flow ($\text{NET} = \text{BUY} - \text{SELL}$). All three groups fluctuate around zero, but their dispersion is time-varying and increases sharply during the COVID-19 window. Retail NET flows show the widest swings and the most prolonged high-variance periods. In contrast, institutional and foreign NET flows are more constrained but still exhibit visible volatility bursts. The figure also reveals periods of co-movement in \emph{volatility} across the groups, even when the daily signs of their NET flows differ. These stylized facts motivate a persistence analysis capable of distinguishing between level effects and the underlying correlation structure. The sustained high volume of retail activity, combined with clustered bursts during shocks and amplified NET variability, suggests the presence of long memory in the order flow, contrasting with the behavior expected from i.i.d.\ noise. %The subsequent analysis quantifies these observations using DFA: we first report static Hurst exponents for each group and flow type (BUY, SELL, and NET), then track regime changes with rolling estimates, and finally validate our findings against shuffled benchmarks.

\begin{figure}[!ht]
\begin{center}
\includegraphics[width=14cm,height=14cm,keepaspectratio]{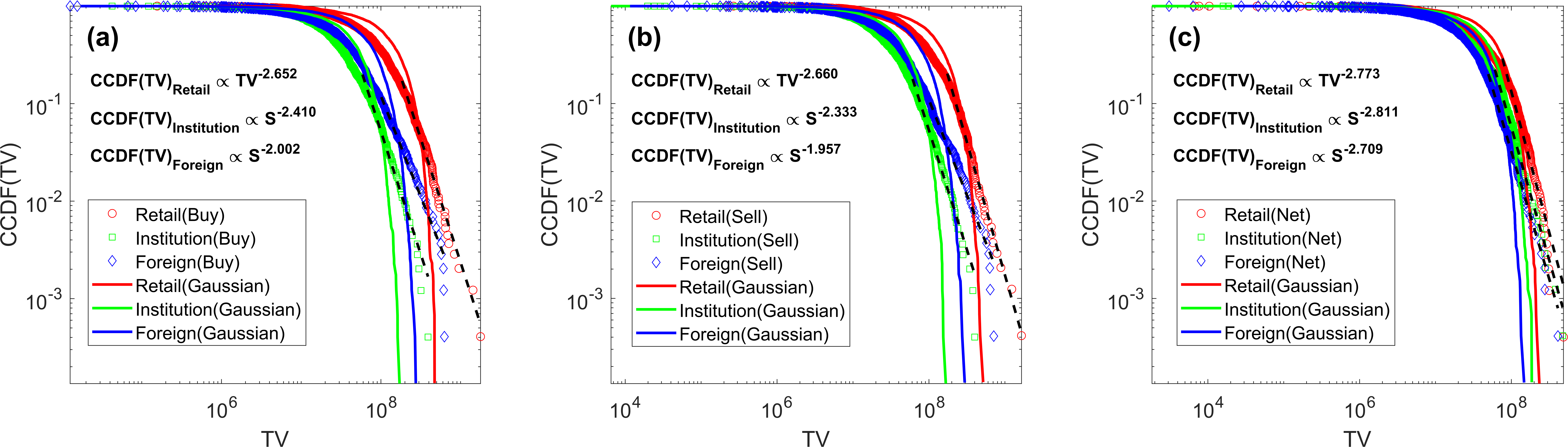}
\end{center}
   \caption{Complementary cumulative distribution functions (CCDFs) of daily trading volumes on a log-log scale. The distributions are shown for (a) BUY, (b) SELL, and (c) NET (BUY$-$SELL). Empirical data for the three investor groups are represented by markers: retail (red), institutional (green), and foreign (blue). For comparison, the solid curves show the CCDF for a Gaussian distribution with the same mean and variance as each empirical series. The dashed black lines indicate power-law fits to the upper tails of the empirical data. In all cases, the tails of the trading-flow distributions are substantially heavier than the Gaussian benchmark, consistent with power-law behavior with exponents ($\alpha$) ranging from approximately 2 to 3.}
  \label{fig:ccdf_tv}
\end{figure}

We begin by testing the Gaussian-i.i.d.\ benchmark that underlies models based on the Efficient Market Hypothesis (EMH). We examine the complementary cumulative distribution function (CCDF) of daily trading amounts for each investor group and flow type, comparing it with a mean-variance-matched Gaussian reference distribution. Figure~\ref{fig:ccdf_tv} shows the resulting log-log CCDFs for BUY, SELL, and NET flows. In all cases, the empirical tails are substantially heavier than the Gaussian benchmark; straight-line segments in the upper tail indicate power-law behavior with exponents ranging from 2 to 3. Panels (a) and (b) show that the Gaussian benchmark is rejected by several orders of magnitude in the upper tail. For the BUY and SELL flows, foreign investors exhibit the heaviest tails (i.e., the smallest exponents, $\approx 2.0$), followed by institutional investors, with retail investors showing the lightest tails. This ordering indicates clear cross-type heterogeneity in the size distribution of daily cash-flow bursts. Panel (c) shows that the net flow ($\text{NET} = \text{BUY} - \text{SELL}$) retains heavy tails but with larger exponents ($\approx 2.7$--$2.8$) and less separation across the groups. This compression is consistent with a partial offset between the BUY and SELL streams: the netting process removes a portion of the extreme-value mass while leaving a non-Gaussian tail. These distributional results strongly reject normality and motivate the use of methods that do not rely on Gaussian innovations. Heavy tails alone, however, do not establish long-range dependence. We therefore turn to DFA to measure temporal persistence directly---first with static Hurst exponents and then with rolling estimates across calm and stress periods.

%\subsection{Temporal correlation: DFA scaling and shuffled benchmark}

Figure~\ref{fig:dfa_both} displays the DFA scaling results. The upper panels (a)--(c) display the outcomes for the original daily flow series, where $\log F(s)$ is plotted against $\log s$ on a log-log scale. The presence of straight lines spanning more than an order of magnitude in scale indicates clear power-law behavior. In the BUY series (Panel a), the fitted slopes are $\alpha=1.359$ for retail, $\alpha=1.116$ for institutions, and $\alpha=1.038$ for foreign investors. The SELL series (Panel b) shows the same ordering with values of $\alpha=1.366$, $1.116$, and $1.018$, respectively. Exponent values above one imply strong persistence, meaning that fluctuations grow with scale faster than would be expected under short-memory dynamics. The similarity between the BUY and SELL exponents within each group shows that the temporal structure is not one-sided; buying and selling flows share a comparable degree of long-range dependence.

For the NET series (Panel c), the slopes compress substantially across all groups: $\alpha=0.674$ for retail, $0.705$ for institutions, and $0.691$ for foreign investors. While these values remain well above the short-memory benchmark ($\alpha=0.5$), they are much smaller than their corresponding BUY/SELL slopes, a finding consistent with partial cancellation between two persistent streams. Importantly, the three groups do not differ significantly in their NET flows; their exponents cluster around similar levels. This suggests that while the investor groups are heterogeneous in their separate buying and selling streams, these differences attenuate once the flows are netted.

The lower panels (d)--(f) report the shuffled benchmarks, which are obtained by randomly permuting the time indices to destroy serial dependence while preserving the marginal distribution. After this process, the slopes collapse toward the theoretical value of $\alpha \simeq 0.5$ for all groups and flow types (e.g., BUY: retail $0.489$, institution $0.466$, foreign $0.626$; SELL: retail $0.478$, institution $0.561$, foreign $0.577$). Small deviations from $0.5$ are expected in finite samples, particularly those with heavy-tailed marginal distributions. Crucially, the clear cross-group ranking observed in the original data largely disappears after shuffling. This contrast confirms that the large exponents reported for the original series arise from their temporal correlation rather than from their marginal distributions.

Taken together, these results establish three key findings. First, in the BUY and SELL flows, retail investors display the strongest persistence, followed by institutional investors, with foreign investors exhibiting the weakest. Second, the netting process attenuates persistence for all groups and causes their exponents to converge, leaving no clear dominance by any single group. Third, the shuffled benchmarks yield slopes consistent with the short-memory value ($\alpha \approx 0.5$), confirming that the persistence observed in the original series is a genuine feature of its temporal structure. These findings support the hypothesis of heterogeneous dynamics in individual BUY and SELL streams but suggest that these group differences diminish when flows are netted, motivating the time-varying analysis with rolling estimates presented in the next subsection.

%\subsection{Rolling persistence and crisis responses}

\begin{figure}[!ht]
\begin{center}
\includegraphics[width=14cm]{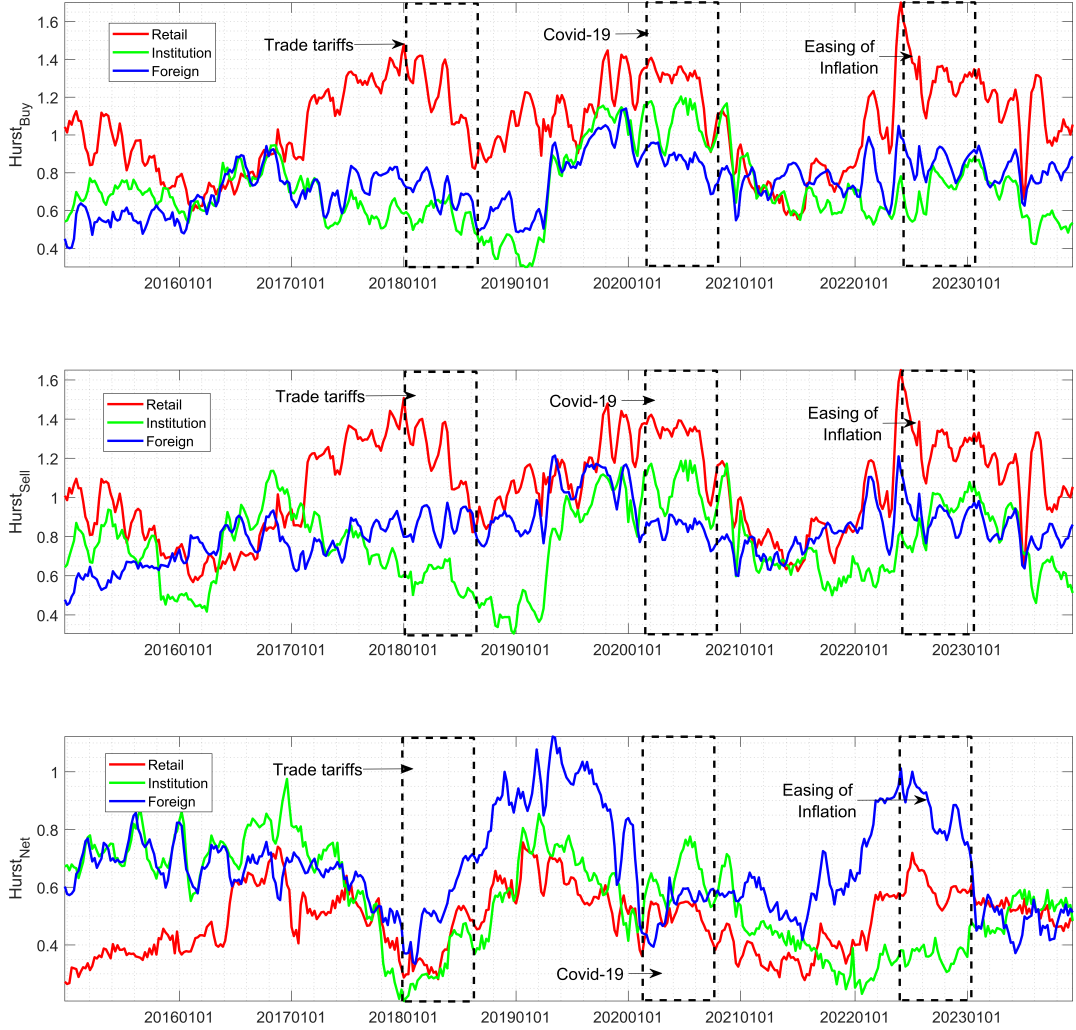}
\end{center}

 \caption{Time evolution of persistence in investor-segregated cash flows, measured by the rolling Hurst exponent ($H_t$). The exponents are computed using a 250-trading-day window with a 5-day step. The panels display the time series of $H_t$ for the BUY (Top), SELL (Middle), and NET (BUY$-$SELL) flows. The different colored lines correspond to the three investor groups: retail (red), institutional (green), and foreign (blue). The vertical dashed boxes highlight the same three periods of market stress shown in Fig.~1: the 2018--2019 trade-tariff tensions, the COVID-19 pendemic, and the 2022--2024 disinflation phase.}
    \label{fig:rollH}
\end{figure}

We next examine how persistence evolves over time and how different investor types respond during periods of market stress. The rolling estimates are computed using a 250-day window with a 5-day step, applying the same admissible scales as in the static DFA. Figure~\ref{fig:rollH} displays the time series of the rolling Hurst exponent (\(H_t\)) for the BUY, SELL, and NET flows. Dashed boxes mark three key periods: the 2018--2019 trade-tariff tensions, the COVID-19 episode (from early 2020), and the 2022--2024 disinflation phase.

The BUY and SELL panels reveal a slow-moving persistence with clear level differences among the groups and visible regime changes around the highlighted periods. The persistence of retail flows is consistently the highest: their \(H_t\) values commonly lie in the 1.1--1.4 range, characterized by broad plateaus and sharp local peaks. Institutional flows exhibit lower levels of persistence, with \(H_t\) values typically between 0.7 and 1.2, while foreign investor flows remain the lowest and most stable. During the trade-tariff period, the persistence of all three groups trends upward, with the increase being most pronounced for retail investors. At the onset of the COVID-19 pandemic, the \(H_t\) for both BUY and SELL flows increases sharply for all groups---a finding consistent with longer execution horizons and increased order splitting under market stress. Afterward, the series tend to mean-revert toward their pre-shock ranges but retain an elevated level of volatility. A second, more transient, period of elevated persistence appears around the 2022--2024 disinflation phase, with the retail series again leading this increase. The similarity between BUY and SELL dynamics within each group indicates that the long-memory structure is not direction-specific; buying and selling flows share a comparable level of persistence.

The results for the NET flows, however, reveal a reversal of the cross-type ranking observed in the BUY and SELL series. Foreign investors now exhibit the highest persistence, with their net flow \(H_t\) values reaching or exceeding unity during the trade-tariff and disinflation periods and remaining elevated for extended durations. Institutional investors are intermediate, while retail investors show the lowest and most volatile NET persistence, with \(H_t\) values frequently falling within the 0.3--0.6 range. This inversion is consistent with the static results and points to a potential mechanism: for retail investors, the persistent buy and sell streams interact like competing processes, causing the netting to cancel a sizable fraction of the long-memory component. Conversely, for foreign investors, buying and selling activities tend to align in direction and time, allowing the netting process to preserve, rather than remove, persistence.

Behavior during crises follows this observed pattern. During the COVID-19 period, while the BUY and SELL \(H_t\) values rise in unison across all groups, their NET \(H_t\) values diverge: the exponent for foreign investors spikes, while that for retail investors remains low or even falls. This indicates that net directional pressure from foreign investors is stronger and more persistent during stress periods, whereas the activity of retail investors remains more balanced between buying and selling. The amplitude of time variation is largest for retail investors in BUY/SELL flows and for foreign investors in NET flows, highlighting distinct adjustment modes by investor type.

Overall, the rolling estimates demonstrate that persistence is not a static property. Instead, it shifts in level across different market regimes and varies significantly by investor type and flow direction. Combined with the evidence of heavy-tailed distributions and the static DFA results, this time-varying analysis strongly favors a heterogeneous view of market activity over a simple short-memory Gaussian benchmark. Specifically, retail flows carry strong and variable long-range dependence on both sides of the market, but this dependence largely cancels upon netting. In contrast, foreign flows maintain a persistent net pressure that survives the netting process, especially within crisis windows.

A crucial question that arises from these findings is whether this observed persistence in investor flows translates into measurable market risk. To investigate this, we perform simple regressions of daily market volatility on the rolling Hurst exponents. This research design allows us to test whether the persistence in investor flows corresponds to systematic variations in market volatility. The analysis provides a direct test of whether these investor-level dynamics can explain market instability beyond the assumptions of short-memory benchmarks. To gauge how investor–level persistence relates to market variability, we run the contemporaneous regression
\begin{equation}
\label{eq:reg_simple}
RV_{t} = \alpha^{(g,a)} + \beta^{(g,a)} H^{(g,a)}_{t} + u_{t},
\end{equation}
where $RV_t$ denotes daily volatility (squared return in the baseline), and $H^{(g,a)}_{t}$ is the rolling DFA estimate for group $g\in\{\text{retail},\text{institution},\text{foreign}\}$ and flow $a\in\{\text{BUY},\text{SELL},\text{NET}\}$ (window $W{=}250$, step $=5$). Table~1 reports slope coefficients and $t$–values from separate OLS fits. 

The association between persistence and volatility is positive in eight of the nine cases and statistically significant in all but one. For the \emph{BUY} and \emph{SELL} flows, the slope coefficient ($\beta$) generally increases across investor types, from retail (0.027--0.030) to institutional (0.047 for BUY, 0.029 for SELL) and foreign (0.079 for BUY, 0.093 for SELL), with t-values ranging from 6 to 14. Thus, days on which an investor group's flow process is more persistent (higher \(H_t\)) are also days of higher market volatility, and this linkage is strongest for foreign investors.

The results for the \emph{NET} flows, however, reveal the same cross-type inversion observed in the rolling analysis. Foreign NET shows a clear positive and significant relationship ($\beta=0.034$, $t=5.213$). Retail NET also has a positive and significant coefficient ($\beta=0.046$, $t=4.718$), despite its low average persistence; when persistence in retail net pressure rises, volatility tends to be higher. In contrast, the coefficient for institutional NET is weak and statistically indistinguishable from zero ($\beta=0.007$, $t=1.043$). This highlights that retail investors, often labeled as ``noise traders,'' convey information about market risk through their persistent flows, whereas institutional investors---conventionally regarded as informed---do not show significant explanatory power in their net flows.

Taken together, these regression results reinforce the heterogeneous-investor narrative suggested by the DFA findings. Retail investors display strong long-memory on each side of the market (BUY and SELL), and this side-specific persistence co-moves with volatility. However, the two sides often compete, causing the persistent buy and sell components to partially cancel in the NET flow. Foreign investors, by contrast, maintain a persistent \emph{net} pressure, and when this net persistence increases, volatility rises most sharply among the three groups.

%We report HAC-robust slope errors, permutation \(p\)-values against the null benchmarks, and block-bootstrap confidence intervals. Robustness checks cover DFA order (1 vs.\ 2), scale grids, window lengths, and light preprocessing (variance stabilization, calendar controls). This network-free design isolates persistence at the investor-type level and yields direct, like-for-like contrasts across BUY, SELL, and NET flows.

\section{Conclusions}

Daily investor-segregated cash flows in the Korean equity market exhibit both fat-tailed distributions and long-range dependence ($H>0.5$ as measured by DFA). In contrast, time-randomized shuffles of the data collapse to $H \approx 0.5$, confirming that the observed correlations are genuinely temporal in origin. Static DFA reveals strong persistence in the BUY and SELL streams of retail and institutional investors; however, these differences largely vanish in the NET flows, where all groups cluster around similar exponent values. Rolling estimates, however, reveal clear \textit{regime shifts}: persistence strengthens during periods of crisis (e.g., the 2018--2019 trade tensions, the 2020 pandemic onset, and the 2022--2024 disinflation phase), and foreign investors, in particular, sustain elevated long memory in their NET flows during these periods.

The regression analysis provides a sharper test of whether investor persistence is tied to market variability. The results show that retail flows, long dismissed as ``noise trading,'' are in fact significantly associated with higher volatility when their persistence rises. By contrast, institutional flows---often regarded as ``informed''---do not exhibit significant explanatory power in their NET flows, while foreign investors exert the strongest and most consistent influence, particularly in their net flows. These findings effectively invert the standard labels assigned to investor types: retail investors contribute meaningful signals related to market risk, institutional investors are comparatively weak in this regard, and foreign investors emerge as the dominant influence during times of stress.

The observed results are consistent with a heterogeneous, multiscale market in which temporal correlations can serve as a useful \textit{order parameter}. Rather than supporting a baseline model of i.i.d.\ returns as suggested by the EMH, our evidence aligns more closely with elements of the \textit{Fractal Market Hypothesis}, as periods of market stress tend to coincide with stronger long-memory regimes.

\begin{acknowledgments}
This study was supported by research fund from Chosun University (2024)
\end{acknowledgments}

\newpage
%\bibliographystyle{apsrev4-2}   % 스타일: Physical Review E 권장 스타일
%\bibliography{refs_v4}    
%\bibliographystyle{unsrt}
%\bibliography{refs_v4}
\bibliographystyle{apsrev}   % APS 권장 스타일
\bibliography{refs_v4}

\newpage

\begin{table}[t]
\centering
\footnotesize
\caption{Regression of daily market volatility on rolling Hurst exponents. The table reports coefficients from contemporaneous OLS regressions of the form $RV_t = \alpha + \beta H_t + u_t$. The dependent variable, $RV_t$, is the daily market volatility (measured as squared return), and the independent variable, $H_t$, is the rolling Hurst exponent for a specific investor group and flow type. The rolling exponents were computed using a window of $W=250$ days and a step size of $s=5$ days. For each regression, the table shows the estimated constant ($\alpha$) and slope coefficient ($\beta$), with corresponding t-values reported in parentheses. The symbols *, **, and *** indicate statistical significance at 10\%, 5\%, and, 1\%, respectively.}
\label{tab:vol_on_hurst_panels}
\setlength{\tabcolsep}{6pt}
\begin{ruledtabular}
\begin{tabular}{lccc}
\multicolumn{4}{c}{\textbf{Panel (a): Retail}}\\
\hline
& BUY & SELL & NET \\
\hline
Constant & $-0.003\,(-0.691)$ & $-0.006\,(-1.354)$ & $0.003\,(0.583)$ \\
Hurst    & $0.027\,(6.035)^{***}$ & $0.030\,(6.869)^{***}$ & $0.046\,(4.718)^{***}$ \\
\hline
\multicolumn{4}{c}{\textbf{Panel (b): Institution}}\\
\hline
& BUY & SELL & NET \\
\hline
Constant & $-0.009\,(-2.275)^{**}$ & $0.003\,(0.821)$  & $0.021\,(5.781)^{***}$ \\
Hurst    & $0.047\,(9.108)^{***}$  & $0.029\,(5.460)^{***}$  & $0.007\,(1.043)$ \\
\hline
\multicolumn{4}{c}{\textbf{Panel (c): Foreign}}\\
\hline& BUY & SELL & NET \\
\hline
Constant & $-0.034\,(-6.491)^{***}$ & $-0.051\,(-9.354)^{***}$ & $0.003\,(0.582)$ \\
Hurst    & $0.079\,(11.464)^{***}$ & $0.093\,(14.127)^{***}$ & $0.034\,(5.213)^{***}$ \\
\end{tabular}
\end{ruledtabular}
\end{table}

\newpage

\newpage

\newpage
% (선택) 한 번에 크기 조절하고 싶으면 매크로를 써도 좋아요.
\newcommand{\SubW}{\linewidth}      % 각 패널 가로폭
\newcommand{\SubH}{0.36\textheight} % 각 패널 세로 상한

\begin{figure}[!ht]
  \centering

  % (a) Original
  \begin{minipage}[t]{\linewidth}
    \centering
    \textbf{Original series}\\[2pt]
    \includegraphics[width=\SubW,height=\SubH,keepaspectratio]{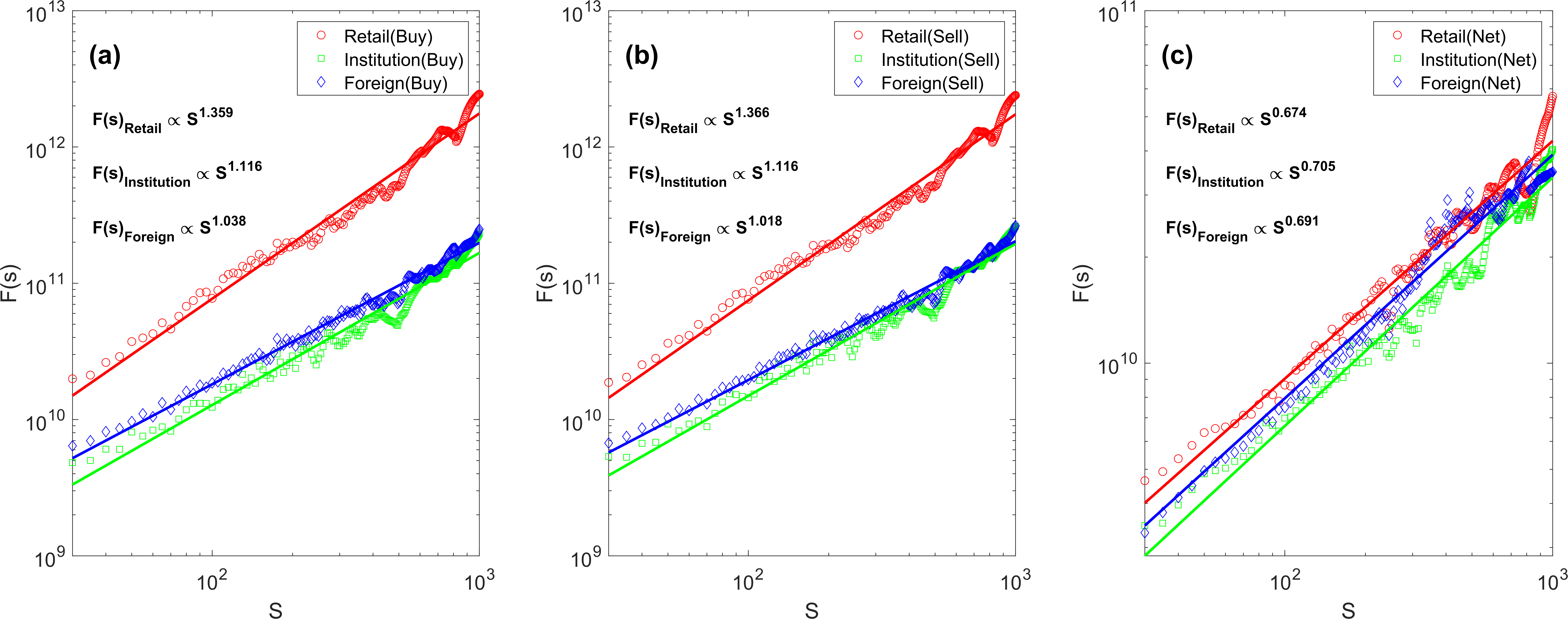}
  \end{minipage}

  \vspace{0.6em}

  % (b) Shuffled
  \begin{minipage}[t]{\linewidth}
    \centering
    \textbf{Shuffled benchmark}\\[2pt]
    \includegraphics[width=\SubW,height=\SubH,keepaspectratio]{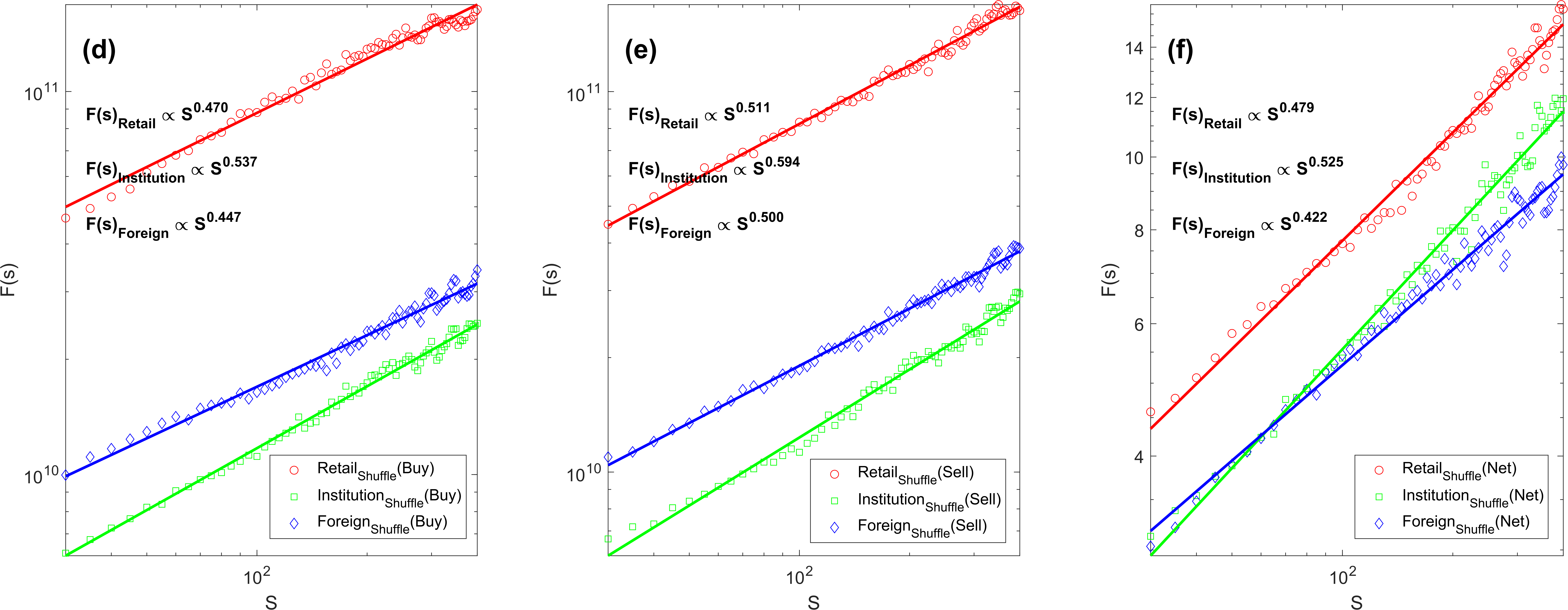}
  \end{minipage}

  \caption{DFA scaling analysis of investor-segregated trading flows from 2015 to 2024. The fluctuation function $F(s)$ is plotted against the time scale $s$ on log-log axes. The top row (panels a--c) displays the results for the original daily time series, while the bottom row (panels d--f) shows the results for their time-shuffled benchmarks. The columns correspond to BUY flows (a, d), SELL flows (b, e), and NET flows (c, f). The solid lines represent the ordinary least squares (OLS) fits to the data, with the resulting scaling exponent $\alpha$ annotated in each panel. The clear linear scaling in the original series (top row) indicates robust long-range dependence. In contrast, the exponents for the shuffled series (bottom row) collapse to the theoretical value of $\alpha \approx 0.5$, confirming that the observed persistence is a genuine feature of the temporal correlations and not an artifact of the data distribution.}
  \label{fig:dfa_both}
\end{figure}

\newpage

\end{document}